\documentclass[12pt,preprint]{aastex}

\shorttitle{Evolutionary stages of M3 variables}
\shortauthors{J. Jurcsik et al.}

\begin{document}

\title{Detection of the evolutionary stages of variables in M3}

\author{J.~Jurcsik\altaffilmark{1}, J.~M. Benk\H o\altaffilmark{1},
G.~\'A.~Bakos\altaffilmark{1,2}, B.~Szeidl\altaffilmark{1}, and
R.~Szab\'o\altaffilmark{1} }
\altaffiltext{1}{Konkoly Observatory of the Hungarian Academy of Sciences, H-1525 Budapest,
PB 67; user@konkoly.hu}
\altaffiltext{2}{Center for Astrophysics, 60 Garden St. Cambridge, MA02138, USA;
gbakos@cfa.harvard.edu}

\begin{abstract}
The large number of variables in M3 provides a unique opportunity to
study an extensive sample of variables with the same apparent distance
modulus. Recent, high accuracy CCD time series of the variables show
that according to their mean magnitudes and light curve shapes, the
variables belong to four separate groups. Comparing the properties of
these groups (magnitudes and periods) with horizontal branch
evolutionary models, we conclude that these samples can be unambiguously 
identified with different stages of the horizontal branch stellar evolution.
Stars close to the zero age horizontal branch (ZAHB)  show Oosterhoff~I
type properties, while the brightest stars have Oosterhoff~II type 
statistics regarding their mean periods and RRab/RRc number ratios.
This finding strengthens the earlier suggestion of \citet{ldz} connecting
the Oosterhoff dichotomy to evolutionary effects, however, it is 
unexpected to find large samples of both of the Oosterhoff type within a
single cluster, which is, moreover, the prototype of the Oosterhoff~I
class globular clusters.
The very slight difference between the Fourier parameters of the
stars (at a given period) in the three fainter samples spanning over about 
0.15 mag range in $M_V$ points to the limitations of any empirical methods which
aim to determine accurate absolute magnitudes of RR Lyrae stars solely from the
Fourier parameters of the light curves.

\end{abstract}


\section{Introduction}

Variable stars in globular clusters are very important objects in
understanding horizontal branch (HB) stellar evolution. They are of the
same age, and the spread in their metallicity, if present at all, is
supposed to be rather small. Their global properties are well known, but
there are still significant uncertainties in tying the basic parameters 
of the globular clusters (distance, age and metallicity) to absolute scales. 
The so called ``second parameter'' which defines the structure of the HB besides 
metallicity is still a matter of debate \citep[see e.g.][]{bella,catelan,bu}.
The systematic differences in the mean periods of RRab stars and in the 
percentage of the overtone variables (Oosterhoff (Oo) dichotomy) cannot 
be simply connected to the HB type or the metallicity of the clusters.
\citet{ldz} suggested an evolutionary explanation of the Oo dichotomy
and the Sandage period shift, arguing that evolution away from the ZAHB can  
explain the observed properties of Oo~II clusters. \citet{lcb} draw a similar 
conclusion from the comparison of M2 (Oo~II) and M3 (Oo~I) but they
found a 2 Gyr age difference between these clusters, also.
The comparison the HB luminosities of different clusters,
which is crucial both in the explanation of the Oo dichotomy and the
period shift, bears, however, significant uncertainties. Thus,
studying stars of different evolutionary stages in a single cluster
may have crucial impact on these studies.  

M3 is one of the most prominent globular clusters with a very
extensive population of RR Lyrae stars (both of RRab and RRc) making
the cluster an ideal target for the investigation of the HB within 
the instability strip. The properties of RR Lyrae stars classify M3 as 
an Oo~I type cluster with a larger population of RRab stars than the 
overtones, and with a 0.561~d mean period of the RRab stars 
\citep[hereafter CC01]{cc01}.
The majority of the RR Lyrae stars have mean magnitudes within a
$\sim0.3$~mag range. The accuracy of photometric data indicates that the 
spread is intrinsic and may be understood in terms of HB evolution.
\citet{k98} mentioned that the three brightest RRab stars might 
already be in an evolved phase of their HB evolution. \citet{cs} have 
drawn attention that these stars fit the period - amplitude (P-A) 
relation of Oo~II clusters. This led them also to conclude that the 
Oo dichotomy is due to evolution and the previously assumed 
period-amplitude-metallicity relation was just an artifact of different 
selection effects. Based on their magnitudes and positions on the 
P-A plot, CC01 separated a group of RRab stars in M3 being probably 
already in the late stage of the HB evolution.

In this Letter we present details on the fine structure of the HB of 
M3 inside the instability strip which strengthen the results of CC01 
and \citet{cs}. The properties of the variables are explained in the 
context of the HB stellar evolution.

\section{Data}

Using all the available photometric V observations of M3 variables
\citep[hereafter BJ03]{k98,c98,cc01,bj} we constructed complete, accurate 
light curves of about 100 RRab and 50 RRc type variables. As our aim 
was to investigate the differences in the shapes of the light curves,
and the magnitude distribution of the variables, only those stars were 
used which were not affected seriously by any type of modulation 
(Blazhko, nonradial, double mode). Constructing the light curves by 
using all the available measurements can help to eliminate any defect 
(distortion) which might be present in any of the observations.

A comparison of the light curves from different observations and reduction
processes \citep{j03} showed that the light curves of the inner variables in 
CC01 have considerably larger scatter and much less reliable mean magnitudes 
than the BJ03 data. For the outer variables, the scatter in the V light 
curves of CC01 is a bit larger than in the BJ03 and \citet{k98} data. 
Consequently, we rely on the correctness of the BJ03 data and to reach the most
accurate light curve shapes we do not simply merge the different data, but if 
necessary, magnitude offsets of the order of 0.01~mag to the different 
measurements are added to match the data of the individual variables to the 
BJ03 light curves. The error of the intensity mean magnitudes of the variables 
used in the this work is typically less than 0.02~mag in BJ03, thus this is the 
typical accuracy of the data the present study is based~on.

The accuracy of the mean magnitudes makes it possible to map the light curve 
shapes and periods (Fourier parameters) in the two dimensional period - magnitude 
plane. Assuming homogeneous composition, this plane is analogous to the 
color-magnitude diagram, as variables with the same magnitudes have nearly equal 
masses within the instability strip, thus differences in their periods are only 
due to their different temperatures (colors).

\section{Results}

Both the RRab and the RRc stars show a relatively wide $\sim 0.3$ magnitude
distribution with two central peaks at around 15.67 and 15.63~mag, a flat 
wide range of brighter stars and a faint tail going down to 15.75~mag. 
The magnitude distribution of the RR Lyrae stars are shown in the top panels 
of Fig.~\ref{fig1}. 
A comparison with HB synthesis results using the evolutionary models of 
\citet{dor} are shown in the bottom panel of Fig.~\ref{fig1}. Uniform mass 
and age distributions of the HB stars  in the 0.70 and $0.60 M_{\sun}$
and $0-100$ Myear intervals are assumed, and the instability strip of 
RR~Lyrae stars are defined as shown in Fig.~\ref{fig4}.
Synthetic HB simulations gave $<M_{HB}>=0.64$ and $\sigma_M=0.02 M_\Sun$
mean mass and mass dispersion values in M3 \citep{cat01}.
However, as different models have different mass distributions on the HB, 
when using \citet{dor} models to synthetize the HB population of M3, other
values of the mean mass and mass dispersion may be obtained.  
Therefore, we tested the synthetic magnitude distribution within the 
instability strip using mean mass and mass dispersion values within 
the $0.62-0.68$ and $0.02-0.03 M_\Sun$ ranges. These simulations led to 
similar results as for uniform mass distribution, with a bit more narrow 
density peak when smaller mass and mass dispersion values were assumed. 

The observed magnitude distributions of RR Lyrae stars and synthetic HB 
results are in good agreement, taking into account the uncertainties of 
both the observations and the models. This global agreement helps in 
identifying the different magnitude groups of RR~Lyrae stars with different 
stages of the HB evolution. 

As there is no clear cut between the magnitudes of the different magnitude 
groups identified in Fig.~\ref{fig1}, the light curve shapes help in deciding 
which group a given star belongs to. In order to check the possible systematic 
differences between the shapes of the light curves at different mean magnitudes, 
we compared the progressions of their Fourier parameters as a function of their 
periods as shown in Fig.~\ref{fig2}.
These plots can be interpreted as changes in the light curve shapes
with decreasing temperature for the four samples which have the same
luminosity, composition, and there should be just a very slight ($<0.03
M_{\Sun}$) dispersion in their masses. The most dramatic differences can 
be seen in the Fourier parameters of the brightest sample, but the other 
brighter group is also slightly shifted, especially in the higher Fourier 
components from the fainter stars. The two faint samples seem to follow 
the same tracks. The larger scatter of the Fourier parameters of the 
faintest sample may be due to observational inaccuracies but it cannot be 
excluded that it reflects intrinsic differences in the light curve shapes 
of these stars. The observed main tracks in Fig.~\ref{fig2} are very similar 
to the predicted behaviour of the Fourier parameters of constant luminosity,
mass and composition models shown by \cite{dorf}. This result strengthens 
the reality that the three brighter groups represent indeed different 
luminosity samples.

The period - Fourier amplitude plots shown in Fig.~\ref{fig2} are the
analogs of the P-A diagram discussed in many papers as a diagnostic tool 
of the Oo type. CC01 separated a sample of 13 RRab stars in M3 which defined 
a long period sequence on the P-A diagram.
According to their observations, 11 of these stars were brighter than the average 
magnitude of the RR Lyrae stars. Our measurements confirm this finding, moreover 
in our data all these stars belong to the brightest sample including an 
additional member of this group; V139.

Fig.~\ref{fig3} shows the period distributions in the four different 
brightness groups, and for comparison, the period distribution in M2 as well.
The period distributions of RRab stars in the four groups show definite
shifts with magnitude.  Both the shortest and longest periods and also 
the distribution of the periods in the third group are very similar to that 
of the two fainter groups, but are shifted by $+0.03$ d. The mean period of
the brightest stars is 0.12 d longer than in the main group.

\citet{j98} explained the Sandage period shift (an increase of the
mean period of RRab stars in globular clusters with decreasing metallicity)
with the inclination of the RRab instability strip. The variables in M3 also 
indicate an inclination of the instability strip (CC01(Fig.~9) and \citet{bakos}). 
The increased luminosities of the two brighter samples account for $\sim0.025$ 
and $\sim0.07$~d longer periods, respectively, according to the linear 
pulsation equation. To explain the observed 0.05 and 0.12~d longer periods 
in these samples, cooler temperatures and/or smaller masses of these stars 
has to be also assumed. 

The period distribution of the faintest stars is the same as that of the most 
populous group. For the smaller sample of the RRc stars, a similar investigation 
of their period distribution is not possible, however, it is worth to note that 
all the six longest period RRc stars belong to the brightest group.

\section{Conclusions}

\begin{figure}[bbb!!!]
\figurenum{1}
\epsscale{.85}
\plotone{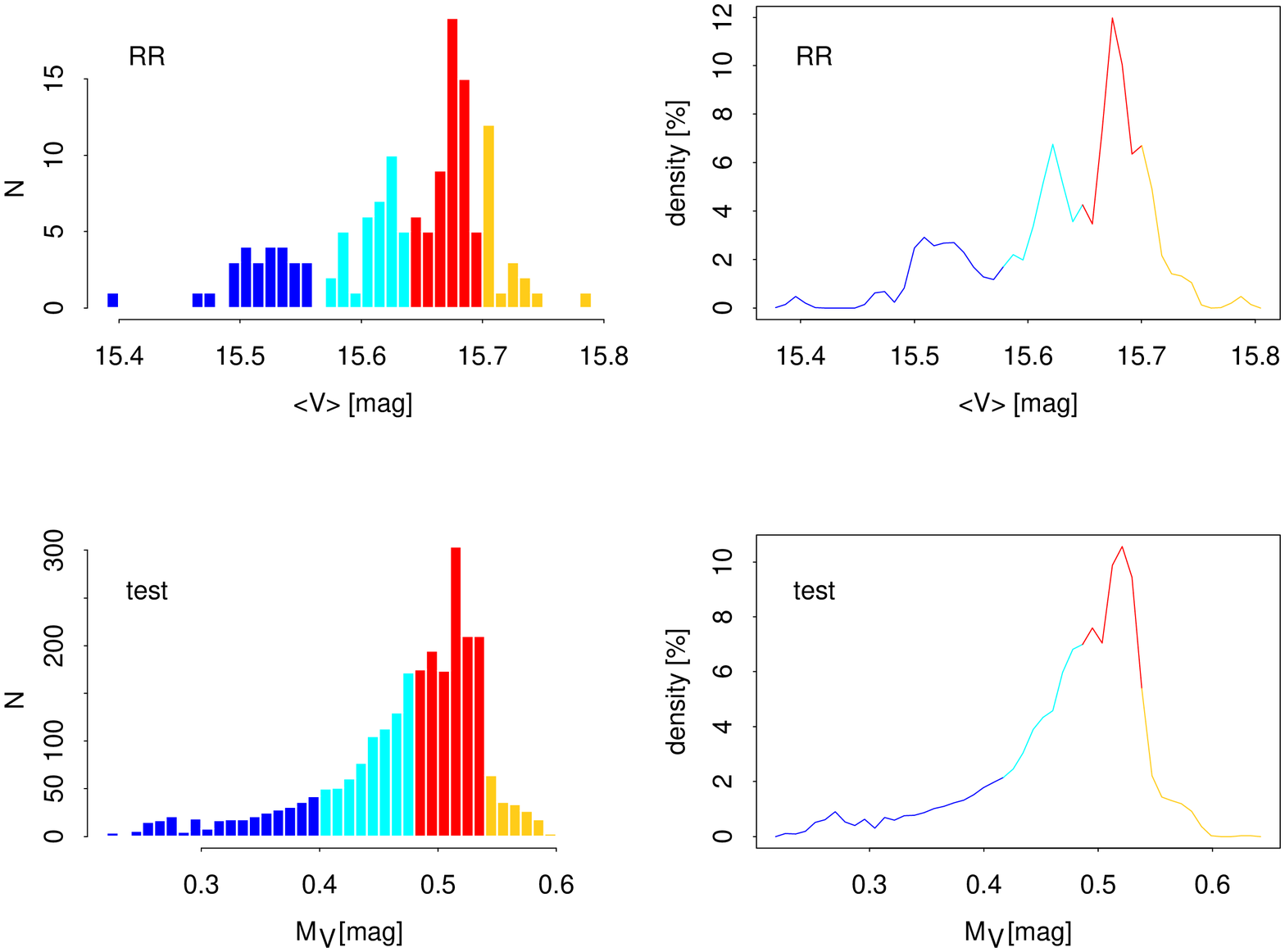}

\caption{Histograms and density functions of the mean V magnitude
distributions of the RR Lyrae stars in M3 (top panels). Results from 
synthetic HB simulations using \citet{dor} evolutionary tracks 
([Fe/H]$=-1.48$~dex, YHB=0.249, [O/Fe]=0.6~dex) for the RR Lyrae 
instability strip (see also Fig.~\ref{fig4}) are shown in the bottom 
panels. Taking into account the facts that a) the observed sample is 
not complete (162 stars with non-modulated, accurate light curves are 
only used), b) the accuracy of the observed magnitudes is 0.02~mag, 
c) HB evolutionary models still bear large uncertainties, -- 
we can conclude that the agreement between observational and model 
results is more than satisfactory.
The lack of model stars at brighter magnitudes can be explained with the 
existence of the 'breathing pulses' which would lengthen the time elapsed 
during the final HB evolutionary stages. Both samples can be divided into 
four different populations: three samples centered at $<V>=15.67$~mag, 
15.62~mag, and 15.53~mag, respectively, and the faintest stars are below 
$<V>=15.7$ mag. The distribution of the brightest sample is rather flat. 
\label{fig1}}
\end{figure}

Based on the comparison of our data with HB evolutionary models, the four 
groups at different mean brightness can be identified with different 
stages of the HB stellar evolution as shown in Fig.~\ref{fig1} and 
Fig.~\ref{fig4}. Table~\ref{tab1} summarizes the observed properties and 
the suggested evolutionary stages of the variables belonging to the four groups. 
The faintest sample (27) can be identified with variables close to their 
ZAHB positions. Most of the stars in our sample (50) are evolving blueward 
on the HB. Stars belonging to the 0.05 mag brighter sample (33) are at 
the hottest part of the blue loops, while the brightest stars (32) are 
in the late, redward phase of their HB evolution. It is interesting to note 
that among the overtone variables most of the stars belong to this group, 
which indicates that in M3, on the average, RRc stars are already in a later 
phase of their HB evolution than the RRab variables.

\begin{figure}[t!]
\epsscale{1.0}
\plotone{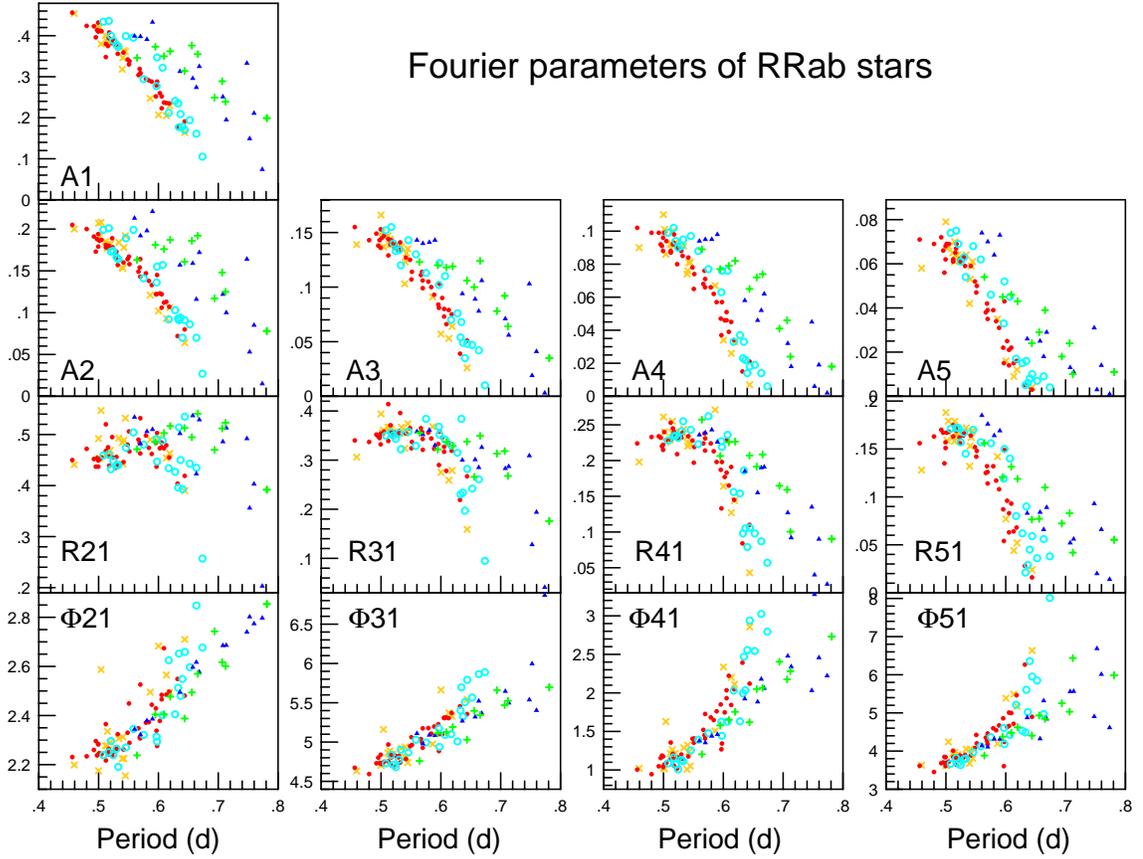}
\figurenum{2}
\caption{Fourier parameters of RRab stars with respect to their pulsation 
period. The four different magnitude groups follow three different
tracks especially in the higher Fourier components. The most dramatic
differences are in the behaviour of the brightest sample (triangles),
whose amplitudes are significantly larger than the amplitudes of the
fainter variables at the same period. The two faintest samples 
(crosses and filled circles) seem to follow the same sequences, the 
larger scatter of the fainter one, though may be real, but larger 
observational uncertainties as a reason cannot be excluded either.
The tracks defined by the fainter samples are very similar to the results 
of the model calculations of \cite{dorf} for the same mass, composition,
and luminosity models at different temperatures.
The observed magnitude range of the three fainter samples is about 0.15~mag.
The similar behaviour of their Fourier parameters indicates serious 
limitations of any empirical methods which aim to determine the absolute 
magnitudes of the variables with hundredths of magnitude accuracy exclusively 
from light curve parameters. For comparison, RRab stars of M2, a typical 
Oo~II cluster with the same metallicity as M3 are also shown by plus signs. 
The overlap of the brightest M3 sample with the M2 variables indicate 
that the brightest stars in M3 share Oo~II properties.
\label{fig2}}
\end{figure}

\begin{figure}[t]
\epsscale{.8}
\plotone{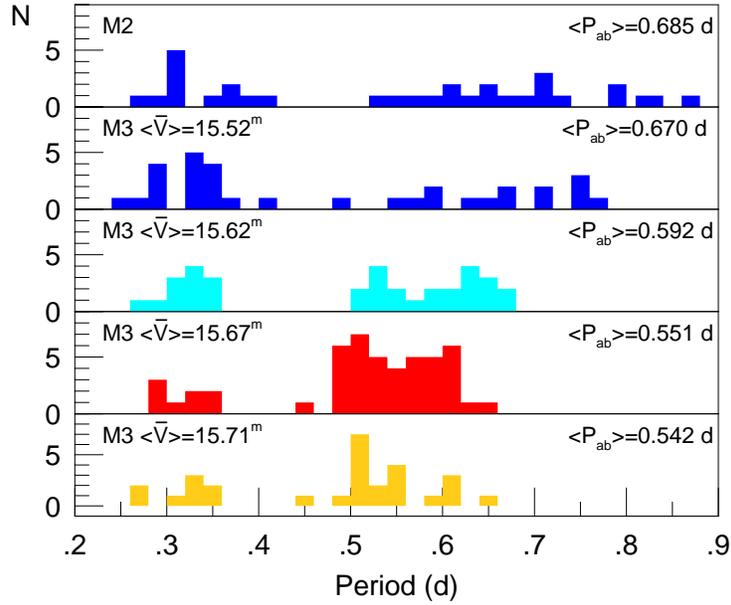}
\figurenum{3}
\caption{Period distributions for the four RR Lyrae samples. The periods of RRab 
stars in the brighter samples are systematically longer than those of the fainter 
variables which are close to the ZAHB. The period distributions of the RRc stars 
do not show any significant differences, however, it may be because of the smaller 
sample sizes. The variable with 0.486~d period in the brightest M3 sample is V70, an 
anomalously long period RRc star. The top panel shows the period distribution in M2.
Although the RRab periods in M2 are even a bit longer than in the brightest group of M3,
both the larger percentage of RRc stars and the longer periods of RRab stars of the
brightest sample resemble to the similar properties of the Oo~type II M2 variables.
\label{fig3}}
\end{figure}

\begin{figure}[tt!!!]
\epsscale{.8}
\plotone{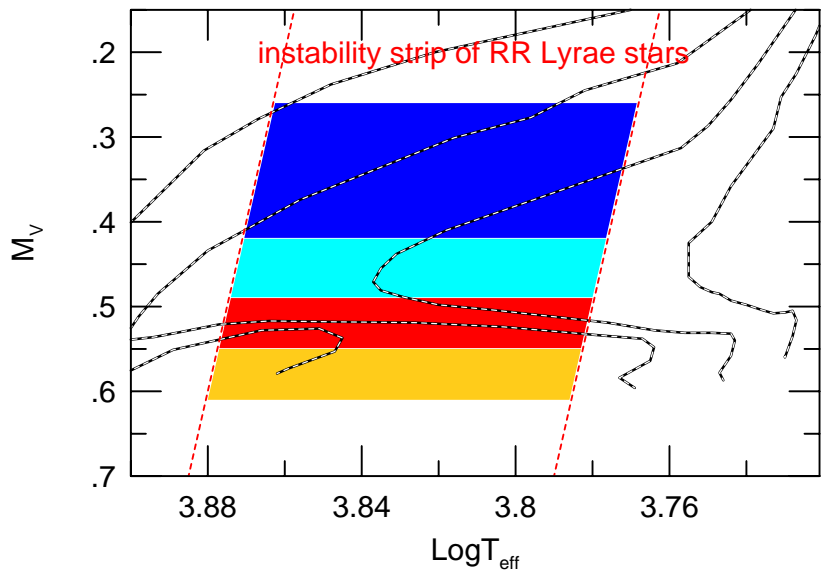}
\figurenum{4}
\caption{Comparison of the magnitude distribution of the four samples with 
HB evolutionary models \citep{dor}. From right to left 0.70, 0.66, 0.64 and 
0.62~$M_\sun$ tracks are shown. The measured magnitudes are shifted by 15.15~mag, 
which agrees within uncertainties with the 15.12~mag apparent distance modulus 
of M3 \citep[2003 upgrade]{h96}. The four samples can be clearly identified with 
the different stages of the HB stellar evolution. The faintest stars are close 
to the ZAHB, the most narrow and most populous sample centered at $M_V=0.51$~mag 
corresponds to stars evolving blueward, the 0.05~mag brighter group is at the 
hottest part of the blue loop, while the brightest stars are already in the late, 
redward stage of the HB evolution. 
\label{fig4}}
\end{figure}

If we compare the mean periods of RRab stars and the percentage of the
RRc variables of the ZAHB and the blueward evolving stars with that of the
brightest, most evolved sample (0.55 and 0.67 d; 22 and 56\%, respectively), 
an Oo~I and an Oo~II population emerge. The Fourier parameters of RRab stars 
in M2, a globular cluster with the same metallicity as M3, but with Oo~II 
properties, are also shown in Fig. 2 using \citet{lca} data. The 11 M2 
variables and the 14 brightest RRab stars in M3 cover similar range in period 
and define the same sequences (V10 in M2 with 0.87 d period is not shown in
Fig.~\ref{fig2}). The large scatter of these tracks is most probably  
intrinsic, and is due to the larger range in the evolutionary status of 
the most evolved HB stars. The period distribution in M2, as shown in 
Fig.~\ref{fig3}, is also similar to that of the brightest M3 sample in
accordance with our conclusion that the brightest stars have Oo~II 
characteristics. This is the first direct evidence of the Oo dichotomy 
in a single cluster with homogeneous metallicity. 

The fact that the Oo dichotomy in M3 can be consistently explained alone by 
evolutionary effects is a strong constraint for its interpretation
and favours to the hysteresis hypothesis originally proposed by \citet{ab}.
The hysteresis, namely, that mode switching from overtone to fundamental 
takes place at a lower temperature than in the opposite direction, may 
account for both the larger percentage of the overtones and the longer 
period of the fundamentals during the late redward phase of the HB.
A similar conclusion has been already drawn by \citet{cs} from the 
comparison of the P-A diagrams of different metallicity clusters, and from 
the existence of 3 anomalously bright and large amplitude RRab stars in M3. 
\citet{ldz} and \citet{lcb} explained the Oo~dichotomy and the period shift 
by HB stellar evolution but their interpretation was drawn from the 
comparison of the properties of different clusters with order of Gyr age 
differences, while the age difference between the Oo~I and Oo~II populations 
in M3 is smaller than 100~Myear.

As a summary, we succeeded in discriminating the different stages of 
HB stellar evolution using accurate light curves and mean magnitudes 
of the variables in M3. A comparison with HB evolutionary models does 
not reveal any significant discrepancy between observations~and model
predictions. This result helps in studying the ``fine structure'' of the
HB and can be the base of a more precise distance estimate of RR Lyrae 
variables. However, the similarity of the light curves (Fourier parameters) 
of variables in a $\sim0.15$~mag magnitude range imposes strong limits on the 
accuracy of any empirical method \citep[e.g.][]{kw} which derives the magnitudes 
of the stars solely form the Fourier parameters of their light curves.

\acknowledgments
This work has been supported by OTKA grant T43504 and T38437.
We would like to thank the stimulating criticism of the anonymous
referee which helped to improve the content of the paper significantly.

\clearpage

\begin{deluxetable}{llllrrc}
\tabletypesize{\scriptsize}
\tablecolumns{7}
\tablewidth{500pt}
\tablecaption{Parameters and evolutionary stages in the four M3 groups
compared with data of M2
\label{tab1}}
\tablehead{
\multicolumn{2}{c}{Mean period / Period range (d)}
&\multicolumn{2}{c}{ $\overline{<V>}$ mag \, s. d. }
&\multicolumn{2}{c}{ No. of stars}
& HB evolutionary phase\\
\colhead{RRab}
&\colhead{RRc}
&\colhead{RRab}
&\colhead{RRc}
&\colhead{RRab}
&\colhead{RRc}
&\colhead{}
}
\startdata
0.685 / $0.528 - 0.876$& 0.333 / $0.273 - 0.420$&  & & 18& 12&  M2\\
\hline
0.670 / $0.560 - 0.774$& 0.336 / $0.251 - 0.486$& 15.533 0.034 & 15.522 0.043& 14& 18& late
redward evolution\\
0.592 / $0.508 - 0.673$& 0.323 / $0.276 - 0.348$& 15.618 0.018 & 15.620 0.010& 22& 11& bluest
part of the blue loop\\
0.551 / $0.456 - 0.644$& 0.316 / $0.284 - 0.353$& 15.671 0.013 & 15.677 0.008& 41&  9& reddest
stage and blueward evolution\\
0.542 / $0.459 - 0.643$& 0.319 / $0.267 - 0.350$& 15.707 0.014 & 15.723 0.029& 19&  8& ZAHB\\
\enddata
\end{deluxetable}

\clearpage

\end{document}